\documentclass[prb,twocolumn,showpacs,superscriptaddress]{revtex4}
\usepackage{amssymb}
\usepackage{graphicx}
\usepackage[tbtags]{amsmath}

\newcommand{\ij}{i\kern -0.08em j}
\input{tcilatex}

\begin{document}

\title{Multiphoton transitions between energy levels in a phase-biased
Cooper-pair box}
\author{V.I. Shnyrkov}
\affiliation{Friedrich Schiller University, Institute of Solid State Physics,
Helmholtzweg 5, D-07743 Jena, Germany}
\affiliation{B.Verkin Institute for Low Temperature Physics and Engineering of the
National Academy of Sciences of Ukraine, 47 Lenin Ave., Kharkov 61103,
Ukraine}
\author{Th.\ Wagner}
\affiliation{Institute for Physical High Technology, P.O. Box 100239, D-07702 Jena,
Germany}
\author{D. Born}
\affiliation{Friedrich Schiller University, Institute of Solid State Physics,
Helmholtzweg 5, D-07743 Jena, Germany}
\author{S.N. Shevchenko}
\affiliation{B.Verkin Institute for Low Temperature Physics and Engineering of the
National Academy of Sciences of Ukraine, 47 Lenin Ave., Kharkov 61103,
Ukraine}
\author{W. Krech}
\affiliation{Friedrich Schiller University, Institute of Solid State Physics,
Helmholtzweg 5, D-07743 Jena, Germany}
\author{A.N. Omelyanchouk}
\affiliation{B.Verkin Institute for Low Temperature Physics and Engineering of the
National Academy of Sciences of Ukraine, 47 Lenin Ave., Kharkov 61103,
Ukraine}
\author{E. Il'ichev}
\affiliation{Institute for Physical High Technology, P.O. Box 100239, D-07702 Jena,
Germany}
\author{H.-G. Meyer}
\affiliation{Institute for Physical High Technology, P.O. Box 100239, D-07702 Jena,
Germany}
\date{\today}

\begin{abstract}
We investigated both theoretically and experimentally dynamic features of a
phase-biased charge qubit consisting of a single-Cooper-pair transistor
closed by a superconducting loop. The effective inductance of the qubit was
probed by a high-quality tank circuit. In the presence of a microwave power,
with a frequency of the order of the qubit energy level separation, an
alteration of the qubit inductance was observed. We demonstrate that this
effect is caused by the redistribution of the qubit level population. The
excitation of the qubit by one-, two-, and three-photon processes was
detected. Quantitative agreement between theory and experimental data was
found.
\end{abstract}

\maketitle

\section{INTRODUCTION}

During the last decade a number of proposals for constructing an artificial
quantum two-level system by making use of mesoscopic Josephson junctions
were implemented~\cite{nak, Friedman00, Wal00, Vion, chio}. Since it was
recognized that these circuits might serve as quantum bits (qubits) for
quantum information devices the field has attracted increased attention.

Basically, two kinds of such devices have been developed, based on the
charge or flux degree of freedom. Here we will consider one realization
only, which is based on the charge degree of freedom. In this quantum system
two charge states differing by $2e$ ($e$ is the electron charge) are mixed
by Josephson tunneling. One example of such a device is the
single-Cooper-pair transistor - two mesoscopic tunnel junctions separated by
a small superconducting island on which the charge can be induced by an
external gate voltage \cite{Tinkham}. The relative energy of the states is
controlled by the gate voltage.

Since the measurement of a quantum system is a very delicate procedure, the
readout sensor is a crucial component of any potential quantum computing
circuit. In order to minimize the exchange of energy between detector and
qubit the control of the reactive component of the output signals has been
proposed and implemented~\cite{ili1}. Such kinds of measurements requires
the proper design of the qubit. For instance a charge qubit can be a
conventional single-Cooper-pair transistor closed by a superconducting loop~%
\cite{fr}. For a certain range of the relationship between effective
Josephson coupling and charge energies $\varepsilon _{J}/E_{C}$ of the
transistor's junctions, this device is effectively a two-level quantum
system with externally controlled parameters~\cite{Tinkham,fr, zor1, krec1}.
Moreover, similar to both the traditional nonhysteretic RF SQUID~\cite{dan,
shnir} and the DC SQUID~\cite{muck}, the phase-biased transistor coupled to
a high-quality radio-frequency tank circuit~\cite{zor3} turns out to be an
ideal parametric converter of charge and flux signals with standard quantum
limit of the energy resolution.

Recently measurements of the energy level separation of a superconducting
charge qubit were reported. The qubit was coupled to a high quality tank
circuit~\cite{born} or non-resonantly to a single mode of the
electromagnetic field of a superconducting on-chip resonator~\cite{wal, bla}%
. Multiphoton transitions between energy levels in superconducting devices
were studied in several articles \cite{nak01, Wallraff03, You, yaponci, Liu}%
; in this work we present both the experimental observation and the
theoretical description of the multiphoton transitions between the ground
and the first excited state in the phase-biased charge qubit \cite%
{Tinkham,fr, zor1, krec1} making use of the impedance measurement technique 
\cite{RandD,ili}.

We begin in Sec. II with a theoretical description of the phase-biased
charge qubit (PBCQ) subjected to a time-dependent gate voltage or magnetic
flux. We calculate the population of the upper level of this effective
two-level system. The expression for the expectation value of the current in
the PBCQ as well as the response of the tank circuit, weakly coupled to a
PBCQ, have also been obtained. In Sec. III we describe the samples
fabrication and the measurement setup. Comparison between theory and
experimental data is discussed in the Sec. IV.

\section{THEORY}

\subsection{Interaction of a phase-biased charge qubit with microwave power}

\begin{figure}[ph]
\includegraphics[width=8cm]{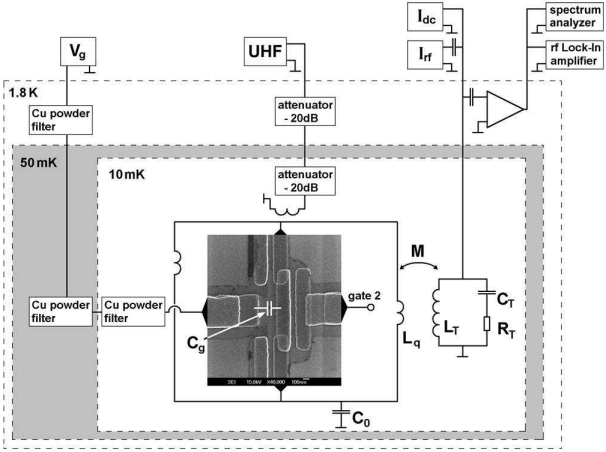}
\caption{Scheme of the PBCQ}
\label{fig1}
\end{figure}

The phase-biased charge qubit (PBCQ), schematically shown in Fig. \ref{fig1}%
, consists of two Josephson junctions closed by a superconducting ring. The
charge $en$ of the island between the junctions is controlled by the gate
voltage $V_{g}$ via the capacitance $C_{g}$, namely by the parameter $%
n_{g}=C_{g}V_{g}/e$; $en_{g}$ is the polarization charge on the island. The
junctions are characterized by the Josephson energies $E_{J1}$, $E_{J2}$ and
the phase differences $\delta _{1}$, $\delta _{2}$. The relevant energy
values are the island's Coulomb energy, $E_{C}=e^{2}/2C_{tot}$, where $%
C_{tot}$\ is the total capacitance of the island, and the effective
Josephson energy $\varepsilon _{J}=\left(
E_{J1}^{2}+E_{J2}^{2}+2E_{J1}E_{J2}\cos \delta \right) ^{1/2}.$ An important
feature of the qubit is that its Josephson energy can be controlled by the
external magnetic flux $\Phi _{e}$ piercing the ring. In this paper, the
ring inductance $L$ is assumed to be small. Consequently, the total phase
difference, $\delta =\delta _{1}+\delta _{2}$, is approximately equal to $%
\delta _{e}=2\pi \Phi _{e}/\Phi _{0}$.

The PBCQ is characterized by the Hamiltonian \cite{Tinkham}:%
\begin{equation}
H=4E_{C}(n-n_{g}/2)^{2}-E_{J1}\cos \delta _{1}-E_{J2}\cos \delta _{2},
\label{Ham_class}
\end{equation}%
which after quantization results in the following Hamiltonian, written in
the representation based on the eigenstates of the operator $\widehat{n}$,
that is in the basis of the charge states $\left\vert n\right\rangle $:%
\begin{eqnarray}
\widehat{H} &=&\sum\limits_{n}4E_{C}(n-n_{g}/2)^{2}\left\vert n\right\rangle
\left\langle n\right\vert +  \label{Ham_quant} \\
&&+\frac{A}{2}\sum\limits_{n}\left( \left\vert n+1\right\rangle \left\langle
n\right\vert +\left\vert n-1\right\rangle \left\langle n\right\vert \right) +
\notag \\
&&+\frac{B}{2i}\sum\limits_{n}\left( \left\vert n+1\right\rangle
\left\langle n\right\vert -\left\vert n-1\right\rangle \left\langle
n\right\vert \right) ,  \notag
\end{eqnarray}%
\begin{equation}
A=-(E_{J1}+E_{J2})\cos \frac{\delta }{2},\text{ }B=-(E_{J1}-E_{J2})\sin 
\frac{\delta }{2}.
\end{equation}%
This Hamiltonian in the two-level approximation can be rewritten in the
basis of the charge states $\{\left\vert 1\right\rangle ,\left\vert
0\right\rangle \}$ \cite{krec1}:

\begin{equation}
\widehat{H}=\frac{A}{2}\widehat{\tau }_{x}+\frac{B}{2}\widehat{\tau }_{y}+%
\frac{C}{2}\widehat{\tau }_{z},  \label{Ham_2_lev}
\end{equation}%
where the irrelevant term containing the unity matrix was omitted;

\begin{equation}
C=4E_{C}(1-n_{g}),
\end{equation}%
and $\widehat{\tau }_{i}$ are the Pauli matrices: $\widehat{\tau }%
_{z}\left\vert 1\right\rangle =\left\vert 1\right\rangle $, $\widehat{\tau }%
_{z}\left\vert 0\right\rangle =-\left\vert 0\right\rangle $.

We consider two possibilities for the excitation of a PBCQ: (a) via gate
voltage: 
\begin{equation}
n_{g}(t)=n_{g}+\widetilde{n}_{g}\sin \omega t,\text{ \ \ }\delta =const,
\label{ng(t)}
\end{equation}%
and (b) via magnetic flux: 
\begin{equation}
\delta (t)=\delta _{0}+\widetilde{\delta }\sin \omega t,\text{ \ \ }%
n_{g}=const.  \label{delta(t)}
\end{equation}

We shall consider first the time-independent case (which we denote by the
subscript \textquotedblleft $0$\textquotedblright ). The eigenstates of the
time-independent Hamiltonian 
\begin{equation}
\widehat{H}_{0}=\frac{A_{0}}{2}\widehat{\tau }_{x}+\frac{B_{0}}{2}\widehat{%
\tau }_{y}+\frac{C_{0}}{2}\widehat{\tau }_{z},
\end{equation}%
denoted by $\{\left\vert -\right\rangle ,\left\vert +\right\rangle \}$, are
related to the charge states $\{\left\vert 0\right\rangle ,\left\vert
1\right\rangle \}$ by the relation:%
\begin{equation}
\left[ 
\begin{array}{c}
\left\vert -\right\rangle \\ 
\left\vert +\right\rangle%
\end{array}%
\right] =\widehat{S}\left[ 
\begin{array}{c}
\left\vert 0\right\rangle \\ 
\left\vert 1\right\rangle%
\end{array}%
\right] \text{.}
\end{equation}%
Here%
\begin{equation}
\widehat{S}=\left[ 
\begin{array}{cc}
\cos \frac{\eta }{2} & e^{i\psi }\sin \frac{\eta }{2} \\ 
-e^{-i\psi }\sin \frac{\eta }{2} & \cos \frac{\eta }{2}%
\end{array}%
\right] ,
\end{equation}%
where the mixing angles $\eta $, $\psi $ are given by: 
\begin{equation}
\sin \eta =\varepsilon _{J}/\Delta E,\text{ \ }\cos \eta =C_{0}/\Delta E,
\label{eta}
\end{equation}%
\begin{equation}
\sin \psi =B_{0}/\varepsilon _{J},\text{ \ }\cos \psi =-A_{0}/\varepsilon
_{J}  \label{psi}
\end{equation}%
with 
\begin{equation}
\varepsilon _{J}=\sqrt{A_{0}^{2}+B_{0}^{2}},
\end{equation}%
\begin{eqnarray}
\Delta E &=&\Delta E(n_{g},\delta _{0})=\sqrt{C_{0}^{2}+\varepsilon _{J}^{2}}%
=  \label{DeltaE} \\
&=&\sqrt{\left[ 4E_{C}(1-n_{g})\right]
^{2}+E_{J1}^{2}+E_{J2}^{2}+2E_{J1}E_{J2}\cos \delta _{0}}.  \notag
\end{eqnarray}%
The diagonalization results in a Hamiltonian in the eigenstate basis: 
\begin{equation}
\widehat{H}_{0}^{\prime }=\widehat{S}^{-1}\widehat{H}_{0}\widehat{S}=\frac{%
\Delta E}{2}\widehat{\sigma }_{z},  \label{Ham_eigen}
\end{equation}%
where we denote the Pauli matrices, which operate in the eigenstate basis,
by $\widehat{\sigma }_{i}$, so that we have $\widehat{\sigma }_{z}\left\vert
+\right\rangle =\left\vert +\right\rangle $, $\widehat{\sigma }%
_{z}\left\vert -\right\rangle =-\left\vert -\right\rangle $.

In order to get the probabilities of the system to be in the eigenstates of
the stationary Hamiltonian $\widehat{H}_{0}$, we rewrite the time-dependent
Hamiltonian $\widehat{H}(t)$ in this basis. This Hamiltonian will be used in
Sec. IV to solve the Bloch-type equation (the master equation) for the
density matrix, whose diagonal elements define these probabilities \cite%
{ShKOK}.

We can split the Hamiltonian $\widehat{H}(t)$ into two parts: 
\begin{equation}
\widehat{H}(t)=\widehat{H}_{0}+\widehat{H}_{1}(t),  \label{H(t)}
\end{equation}%
which gives: 
\begin{equation}
\widehat{H}^{\prime }(t)=\widehat{S}^{-1}\widehat{H}(t)\widehat{S}=\frac{%
\Delta E}{2}\widehat{\sigma }_{z}+\widehat{S}^{-1}\widehat{H}_{1}(t)\widehat{%
S}.  \label{Ham_in_eigenstates}
\end{equation}

We consider first case (a), where the gate voltage is the time-dependent
parameter as in Eq.(\ref{ng(t)}). From Eq. (\ref{Ham_in_eigenstates}) it
follows:%
\begin{eqnarray}
\widehat{H}_{a}^{\prime }(t) &=&\frac{\Delta E}{2}\widehat{\sigma }%
_{z}-2E_{C}\widetilde{n}_{g}\sin \omega t\times  \label{H_a} \\
&&\times \left[ \cos \eta \cdot \widehat{\sigma }_{z}-\sin \eta \sin \psi
\cdot \widehat{\sigma }_{_{y}}+\sin \eta \cos \psi \cdot \widehat{\sigma }%
_{x}\right] .  \notag
\end{eqnarray}

We consider now case (b), where the magnetic flux is the time-dependent
parameter, as in Eq. (\ref{delta(t)}). In this case the time-dependent
Hamiltonian (\ref{H(t)}) can be rewritten by expanding the quantities $A$
and $B$ in a Fourier series. However, we are interested only in small
time-dependent perturbations and for the description of the experimental
results we restrict ourselves to this case, when $\delta _{0}=\pi ,$ \ $%
\widetilde{\delta }/2\ll \pi $. In the second approximation in $\widetilde{%
\delta }$ the Hamiltonian can be written as%
\begin{eqnarray}
\widehat{H}_{b}^{\prime }(t) &=&\frac{\Delta E}{2}\widehat{\sigma }%
_{z}+\left( E_{J1}+E_{J2}\right) \frac{\widetilde{\delta }}{4}\sin \omega
t\cdot \widehat{\sigma }_{x}+  \label{Ham_2nd_approx_d1} \\
&&+\left( E_{J1}-E_{J2}\right) \frac{\widetilde{\delta }^{2}}{32}\cos
2\omega t\cdot \left\{ \frac{\varepsilon _{J}}{\Delta E}\widehat{\sigma }%
_{z}-\frac{C}{\Delta E}\widehat{\sigma }_{y}\right\} .  \notag
\end{eqnarray}

\subsection{Qubit-tank circuit arrangement}

The current in a qubit ring is probed by the tank circuit, which is weakly
coupled through a mutual inductance $M$ to the PBCQ \cite{RandD,ili}. The
PBCQ is characterized by the inductance $L_{PBCQ}=L+L_{J}$, where $L$ is the
ring's inductance and $L_{J}$\ is an inductance defined by $%
L_{J}^{-1}=\left( 2e/\hbar \right) \partial I/\partial \delta $. The effect
of this inductance on the tank circuit can be represented by an effective
inductance: $L_{T}\rightarrow L_{eff}=L_{T}+M^{2}/L_{PBCQ}$ (where $M$ is
the mutual inductance) \cite{RandD}.

The experimentally measurable value is the phase shift between the voltage
and current in the tank circuit $\alpha $. The expression for the phase
shift at the resonant frequency $\omega _{T}=1/\sqrt{L_{T}C_{T}}$ is (see
e.g. in Ref. [26]):%
\begin{equation}
\tan \alpha \simeq k^{2}Q\cdot \frac{L}{L_{J}},  \label{alpha_class}
\end{equation}%
Here $Q^{-1}=\omega _{T}C_{T}R_{T}$, $k^{2}=M^{2}/(L\cdot L_{T})$, and we
have neglected the ring's inductance $L$ in the denominator. Thus the phase
shift $\alpha $ is defined by $L_{J}$, i.e. by the current-phase relation $%
I(\delta )$. This can be used to define the current-phase dependence in
Josephson junctions \cite{RandD,gol}. In the quantum case the current is
equal to the expectation value of the current operator: $I=\left\langle 
\widehat{I}\right\rangle $. For our system we have \cite{Tinkham,krec1}: $%
\widehat{I}=-I_{0}\widehat{\sigma }_{z}$, where the ground state current is%
\begin{equation}
I_{0}=\frac{e}{\hbar }E_{J1}E_{J2}\frac{\sin \delta }{\Delta E}.
\end{equation}%
Thus, $I=I_{0}Z$, where $Z=\left\langle \widehat{\sigma }_{z}\right\rangle
=Sp\left( \widehat{\rho }\widehat{\sigma }_{z}\right) $, and $\widehat{\rho }
$ is the reduced density matrix. This means that the current flows with the
probability $P_{+}$ in one direction and with the probability $P_{-}=1-P_{+}$
in the other direction; and the introduced value $Z$ is equal to $Z=1-2P_{+}$%
. Thus from Eq. (\ref{alpha_class}) for the time-averaged phase shift $%
\alpha $ we have: 
\begin{equation}
\tan \alpha \simeq k^{2}Q\cdot \frac{2e}{\hbar }L\cdot \left\{ \frac{%
\partial I_{0}}{\partial \delta }\cdot \overline{Z}+I_{0}\cdot \frac{%
\partial \overline{Z}}{\partial \delta }\right\} ,  \label{alpha}
\end{equation}%
where the bar stands for the time-averaging. We note that for a weakly
driven system the function $\overline{P}_{+}(\omega ,\delta ,n_{g})$ has the
maxima (resonant peaks) at $\Delta E(\delta ,n_{g})=K\cdot \hbar \omega $ ($%
K $ is an integer) and, consequently, its derivative (see Eq. (\ref{alpha}))
has hyperbolic-like behavior.

In the particular case, when $\delta =\pi $ we have $I_{0}=0$ and from Eq. (%
\ref{alpha}) it follows:%
\begin{equation}
\tan \alpha \simeq -\lambda \cdot \left( \frac{\Delta E}{E_{C}}\right)
^{-1}\cdot (1-2\overline{P_{+}}),  \label{alpha_at_d=pi}
\end{equation}%
\begin{equation}
\lambda =k^{2}Q\cdot \frac{2e^{2}LE_{C}}{\hbar ^{2}}\cdot \frac{E_{J1}E_{J2}%
}{E_{C}^{2}}.  \label{lambda}
\end{equation}%
This means that the dependence of $\alpha $ on $n_{g}$ at $\delta =\pi $
contains resonances at $\Delta E(n_{g})=K\cdot \hbar \omega $.

\section{Sample preparation and measurements}

A SEM image of the gradiometer-type charge qubit's core with two closely
spaced mesoscopic Josephson junctions is shown along with the electrical
circuitry in Fig.~\ref{fig1}. The junction areas are slightly larger than 100%
$\times $100 nm$^{2}$ leading to critical currents $I_{C}$ of 50--100 nA,
which can be estimated from the measured tunnel resistances. The Cooper-box
charge (on the island between the Josephson junctions) can be continuously
varied by the gate voltage. Both the junctions and the island are imbedded
in a macroscopic (0.5$\times $1mm$^{2}$) superconducting
gradiometer-type-loop, which was done in order to minimize vibration and
magnetic noise. The single-Cooper-pair transistor and the loop were
fabricated by e-beam lithography and shadow evaporation of aluminum. One
loop of the gradiometer is inductively coupled by a flip-chip configuration
to the niobium high quality tank coil.

We study the \textquotedblleft qubit + tank\textquotedblright\ impedance as
a function of the polarization charge $en_{g}$ and the phase difference $%
\delta $, by making use of a well-known impedance measurement technique. The
tank circuit is driven by an rf current $I_{rf}$ of frequency $\omega _{T}$
close to the resonance frequency of the tank circuit. The phase difference $%
\alpha $ of the tank voltage (with respect to the phase of the applied
current $I_{rf}$) is measured as a function of the gate voltage $%
V_{g}=V_{g}(n_{g})$ and of the external magnetic flux $\Phi _{e}=\Phi
_{e}(\delta )$. These measurements show a shift in the resonant frequency of
the \textquotedblleft qubit + tank\textquotedblright\ arrangement due to a
change in the effective inductance of the sample. The tank voltage was
sequentially amplified by means of a cryogenic rf preamplifier, a room
temperature amplifier, and further relevant standard electronics.

The measurements were carried out in a dilution-type refrigerator at a
nominal temperature of about 10 mK. In order to minimize the noise level
inherent in the total gate voltage, we equipped the transistor's gate line
(a ThermoCoax between 2 K and 10 mK) with conventional low-pass RC and
microwave copper powder filters (Fig.~\ref{fig1}). For an efficient
thermalization of the charge gate, three microwave filters were mounted on
different low-temperature plates (2 K, 50 mK, and 10 mK) of the
refrigerator. The power attenuation of these 10 cm long filters was
determined as a function of the frequency (up to 45 GHz) at room
temperature. From the measurement results, we concluded that the total
attenuation of this line was more than 80 dB in the GHz range. A
high-attenuation ThermoCoax line along with two (on the 2 K and 10 mK)
cooled commercial 20 dB-attenuators were used for applying a microwave power
(\textquotedblleft UHF gate\textquotedblright ) to the sample. This
microwave line is coupled inductively to the Pb-shield resonance cavity with
the qubit inside. At the microwave frequency the current fed into the Pb
shield is amplified by the quality factor of the resonator, producing an
electromagnetic field. In our measurements special care was taken to avoid
the magnetic coupling between the microwave line and the qubit: (i) the
qubit sample was placed across a Pb resonator for maximum electric (E)-field
interaction and (ii) the gradiometer-type topology of the qubit circuit
prevents the sample from the interaction via mutual inductance with the
microwave line. As a further evidence for E-coupling via the length of the
Al thin film charge-gate electrode, only the noise-like output signal on
microwave power was obtained with sample after mechanical break of this thin
film gate line.

By passing a dc bias current through the tank coil (Fig.~\ref{fig1}) we
could simply control the flux-induced currents circulating in the qubit
ring, because of the mutual inductance between qubit and tank. For the tank,
we prepared square-shaped Nb pancake coils on oxidized Si substrates \cite%
{tor}. For flexibility, only the coil was made lithographically. We use an
external capacitance $C_{T}$ to be able to change the resonance frequency of
the tank which, in this particular case was 28.9 MHz. The tank circuit was
coupled by a 30 cm long piece of two-wire line to the cold HEMT-amplifier.
Changes of the phase of the tank voltage oscillations due to variations of
the Josephson inductance of the sample were measured by means of an
averaging procedure: every measurement point was taken (with a time constant
of 0.1 ms) 50 times and averaged. Cryogenic 
$\mu$%
-metal and superconducting shields protected the sample against external
magnetic and electric noises. However, we could not take any special action
to avoid the drift of electrostatic carriers within the substrate, so the
1/f noise due to background charge motion is not completely negligible in
our experiments.

\section{Results and discussion}

\subsection{Analysis of the experimental results}

\begin{figure}[h]
\includegraphics[width=4.5cm]{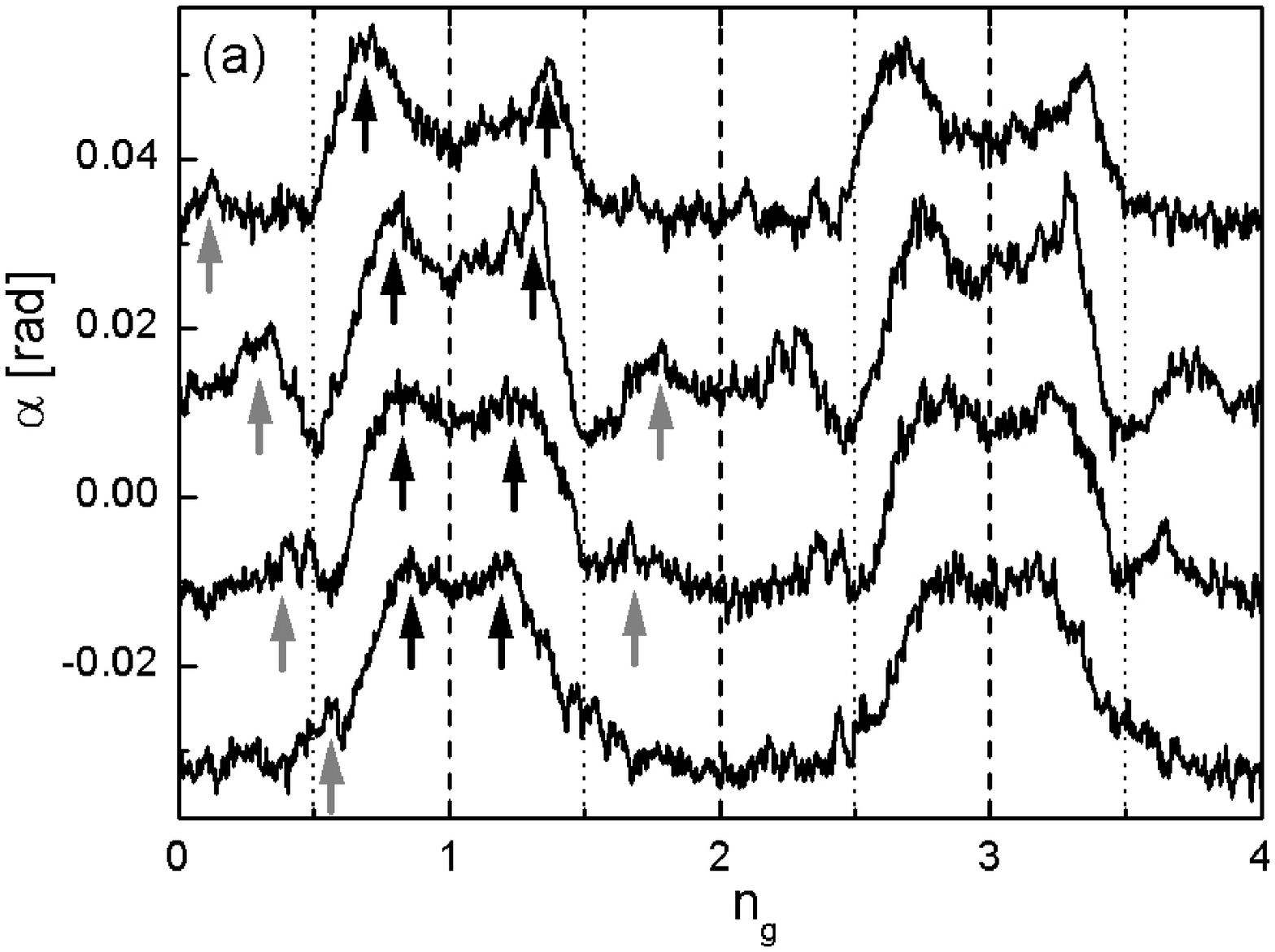}%
\includegraphics[width=4.5cm]{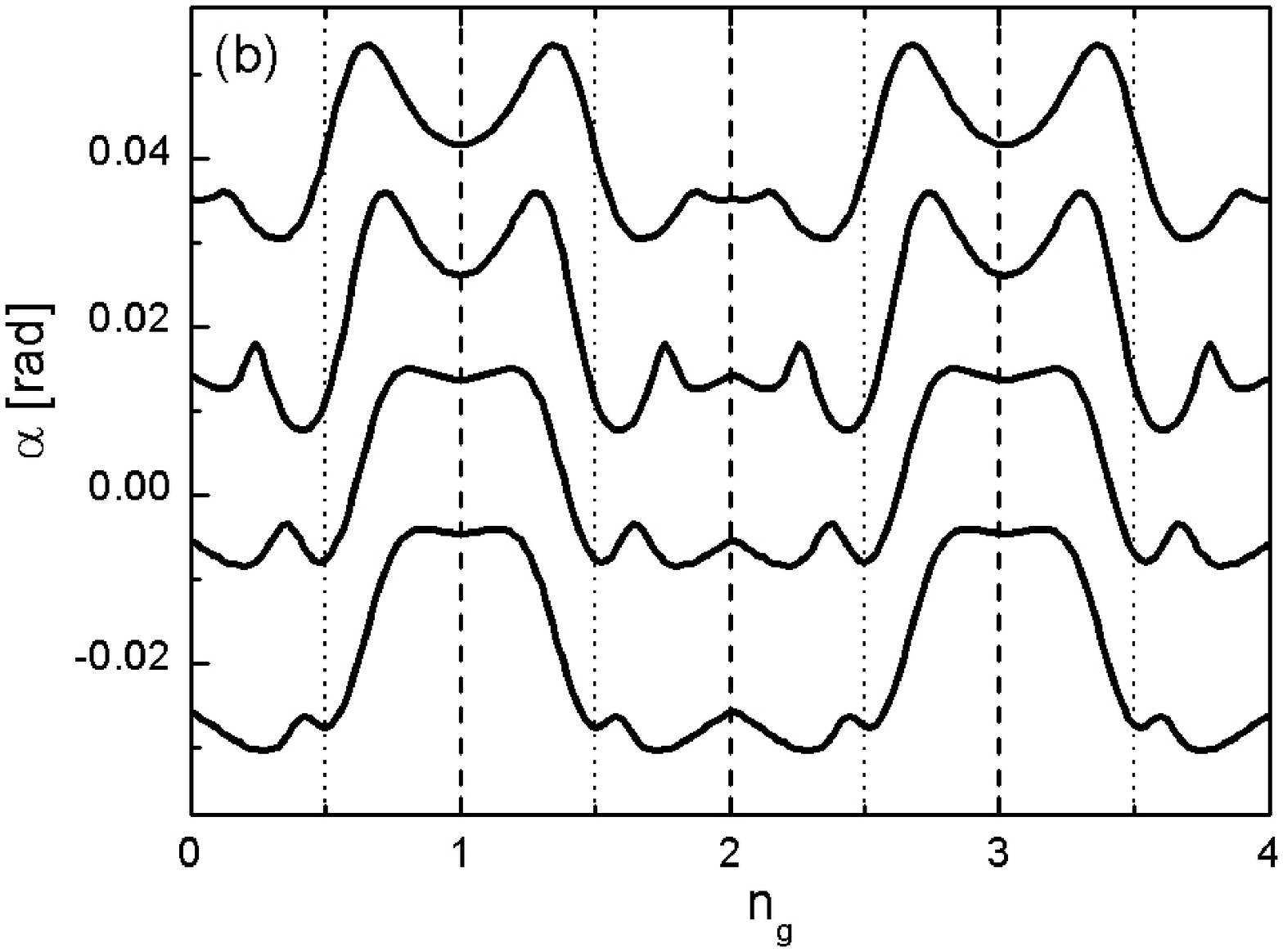}
\caption{Resonant excitation of the PBCQ: dependence of the phase shift $%
\protect\alpha $ on the time-independent part of the dimensionless gate
voltage $n_{g}$ at $\protect\delta =\protect\pi $. The curves correspond to
a fixed power of excitation in experiment (a) and an amplitude $\widetilde{n}%
_{g}\simeq 0.3$\ in theory (b). The varied parameter is the frequency $%
\protect\omega /2\protect\pi $, which from the bottom to top curves is: 6.5,
7.1, 8.1, 9.1 GHz. Upper curves are shifted vertically for clarity. Black
(grey) arrows show the one- (two-) photon resonances.}
\label{Fig2}
\end{figure}

\begin{figure}[h]
\includegraphics[width=4.5cm]{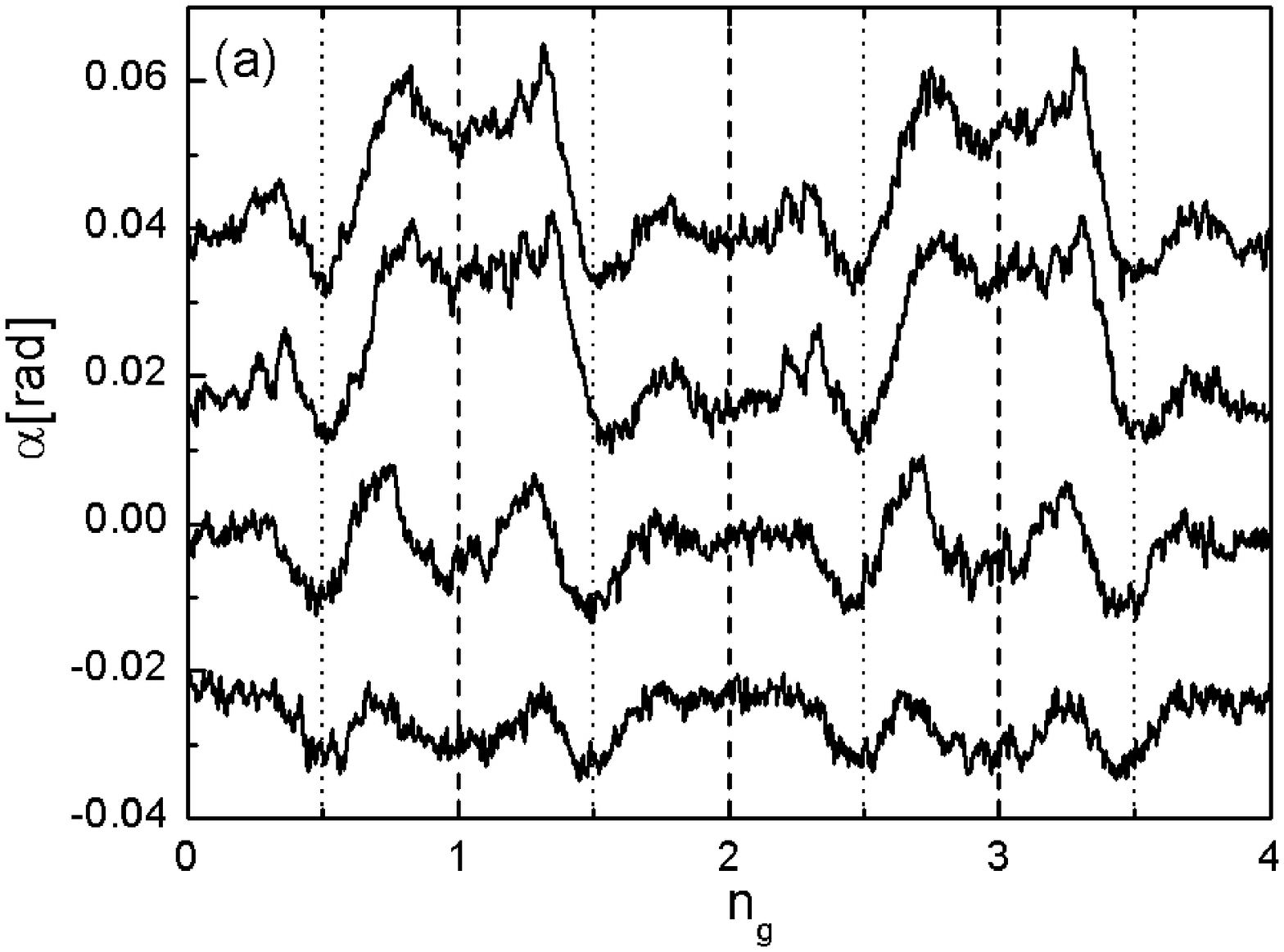}%
\includegraphics[width=4.5cm]{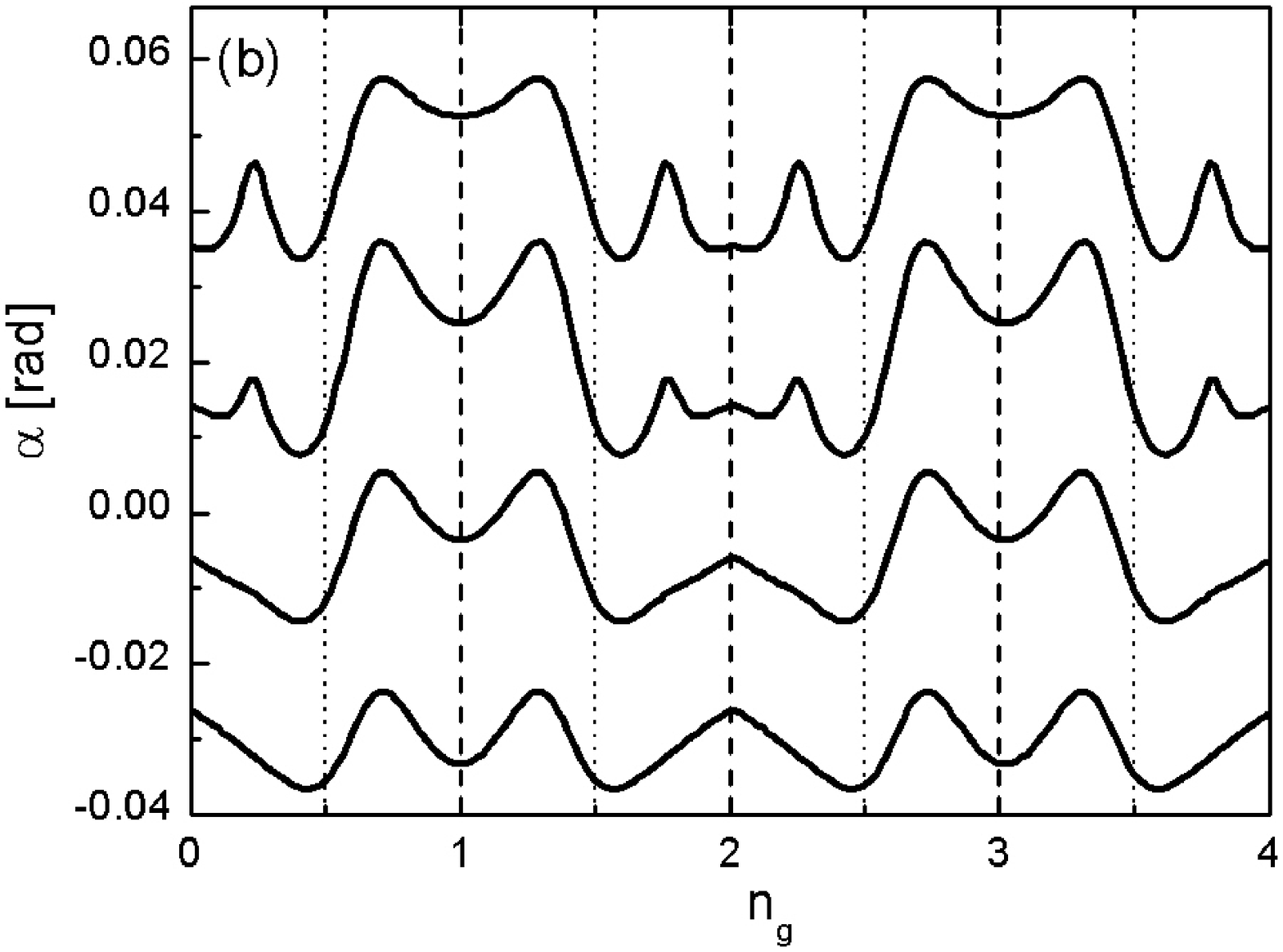}
\caption{The same as in Fig. 2 but at a fixed frequency $\protect\omega /2%
\protect\pi =8.2$ GHz with the varied parameter: in experiment (a) being
power of excitation (from bottom to top: -75, -63, -49, -42 (dB)) and in
theory (b) being amplitude $\widetilde{n}_{g}$ (from bottom to top: 0.1,
0.2, 0.3, 0.4). Upper curves are shifted.}
\label{Fig3}
\end{figure}

The dependence of the phase shift $\alpha $ on the gate voltage is shown in
Figs.~\ref{Fig2}a and~\ref{Fig3}a. In the first set of experiments, we used
different frequencies of the microwave excitation at a nominally fixed power
(see Fig.~\ref{Fig2}a). In a similar manner to the results reported before 
\cite{born} the $\alpha (n_{g})$ dependence exhibits clear peaks. Their
positions depend on the frequency of excitation. Recently it was shown, that
the peaks are due to resonant excitations of the system from ground to upper
states \cite{born}. In the obtained dependencies a second set of the peaks
is clearly seen (the grey arrows in Fig.~\ref{Fig2}). These
\textquotedblleft additional\textquotedblright\ peaks would be due to
two-photons excitation. In order to clarify this issue we fixed the
frequency of the excitation and measured $\alpha (n_{g})$ for different
microwave powers. Indeed, as was expected, the \textquotedblleft
additional\textquotedblright\ peak structure becomes clearer for higher
powers (see Fig.~\ref{Fig3}a). Moreover, an additional structure appears in
the $\alpha (\Phi _{e})$ dependence as well (see Fig.~\ref{Fig5}a).

\begin{figure}[h]
\includegraphics[width=7cm]{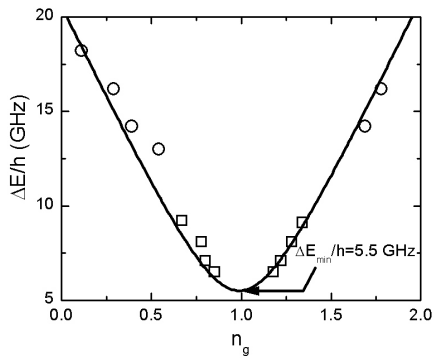}
\caption{Energy level separation $\Delta E$ as a function of $n_{g}$ at $%
\protect\delta =\protect\pi $. Squares and circles correspond to one- and
two- photon resonances; solid line is a fit with $E_{C}/h=5$ GHz and $\Delta
E_{\min }/h=(E_{J1}-E_{J2})/h=5.5$ GHz.}
\label{gap}
\end{figure}

Let us extract the value of the minimum energy level separation $\Delta
E_{\min }=E_{J1}-E_{J2}$\ from the experimental results. In order to do that
we define the position of the resonances, marked with the arrows in Fig. \ref%
{Fig2}a: they correspond to $\Delta E(n_{g})=K\cdot \hbar \omega $. We put
these points in the $n_{g}-\Delta E$-plane: see Fig. \ref{gap}, where the
circles and squares correspond to one- and two- photon resonances,
respectively, for which $K=1,2$. The fitting of these data with the
expression 
\begin{equation}
\Delta E(n_{g},\delta =\pi )=\sqrt{\left[ 4E_{c}\left( 1-n_{g}\right) \right]
^{2}+(E_{J1}-E_{J2})^{2}}
\end{equation}%
allows us to estimate both $E_{C}$ and $\Delta E_{\min }=\Delta
E(n_{g}=1,\delta =\pi )=E_{J1}-E_{J2}$ (see Fig. \ref{gap}).

\subsection{Numerical calculations}

In this subsection we present the results of the quantitative description of
the system. In order to obtain the dependence of the tank voltage phase
shift $\alpha $ on the system's parameters we made use of Eqs. (\ref{alpha})
and (\ref{alpha_at_d=pi}). The probability that the upper level is occupied, 
$P_{+}(t)$, was obtained from the solution of the master equation for the
density matrix as described in Ref. \cite{ShKOK}. In order to take into
account the relaxation and dephasing processes the corresponding rates $%
\Gamma _{relax}$ and $\Gamma _{\phi }$ are included in the master equation
phenomenologically \cite{Blum}. We note that the numerical solution of the
master equation is the general approach used to describe the nonlinear
dynamic behavior of a two-level system subjected to an external field of
arbitrary amplitude and frequency (see in \cite{ShKOK} and references
therein). When the amplitude of the field is small and its frequency is
close to the energy level separation devided by an integer, the analytical
consideration, known as the rotating-wave approximation, can be applied to
the description of the multiphoton transitions (see e.g. \cite{CohTan}). The
latter approach was developed for the description of the PBCQ both
analytically \cite{Krech05} and numerically \cite{Amft}.

In order to unify expressions (\ref{H_a}) and (\ref{Ham_2nd_approx_d1}) and
to get the equations needed for numerical calculations, we write down the
Hamiltonian $\widehat{H}^{\prime }$ as follows: 
\begin{equation}
\widehat{H}^{\prime }\equiv \frac{R}{2}\widehat{\sigma }_{x}+\frac{S}{2}%
\widehat{\sigma }_{y}+\frac{T}{2}\widehat{\sigma }_{z}.
\end{equation}%
Consequently, the evolution of the reduced density matrix $\widehat{\rho }$
taken in the form 
\begin{equation}
\widehat{\rho }=\frac{1}{2}\left[ 
\begin{array}{cc}
1+Z & X-iY \\ 
X+iY & 1-Z%
\end{array}%
\right] \text{,}
\end{equation}%
is described by the master equation in the form of Bloch equations (see in 
\cite{ShKOK}, \cite{Blum}): 
\begin{align}
\frac{dX}{dt}& =\frac{S}{\hbar }Z-\frac{T}{\hbar }Y-\Gamma _{\phi }X,
\label{eq1} \\
\frac{dY}{dt}& =-\frac{R}{\hbar }Z+\frac{T}{\hbar }X-\Gamma _{\phi }Y,
\label{eq2} \\
\frac{dZ}{dt}& =\frac{R}{\hbar }Y-\frac{S}{\hbar }X-\Gamma _{relax}\left(
Z-Z(0)\right) .  \label{eq3}
\end{align}%
From these equations we get $Z(t)$ which defines the occupation probability
of the upper level $\left\vert +\right\rangle $, $P_{+}(t)=\rho _{22}(t)=%
\frac{1}{2}(1-Z(t))$. We choose the initial condition to be $X(0)=Y(0)=0$, $%
Z(0)=1$, which corresponds to the system being in the ground state $%
\left\vert -\right\rangle $.

Quantitative analysis with Eqs. (\ref{Ham_2nd_approx_d1}) and (\ref{eq1})-(%
\ref{eq3}) has shown that the case when the magnetic flux is time-dependent
(Eq. (\ref{delta(t)})) is not consistent with the experimental results
presented in this work. By that we have confirmed the argument, presented in
Sec. III, that the qubit mainly is not excited via the magnetic flux, but
rather via the gate voltage. So in what follows we will consider the case
when the time-dependent parameter is the gate voltage (Eq. (\ref{ng(t)})).
Then the system is described by the Hamiltonian of Eq. (\ref{H_a}).

We consider first the dependence of the phase shift $\alpha $ on $n_{g}$ at $%
\delta =\pi $. The results of the numerical calculation are shown in Figs. %
\ref{Fig2}b and \ref{Fig3}b. In order to fit the experimental curves, by
making use of Eqs. (\ref{alpha_at_d=pi}) and (\ref{eq1})-(\ref{eq3}), we
have taken: $\Gamma _{\phi }/(E_{C}/h)\sim 0.3$ and $\Gamma
_{relax}/(E_{C}/h)\sim 0.05$ (which corresponds to the following decoherence
and relaxation times: $T_{\phi }=\Gamma _{\phi }^{-1}\simeq 0.7ns$ and $%
T_{relax}=\Gamma _{relax}^{-1}\simeq 4ns$) and $\lambda =0.1$. This value of 
$\lambda $ is in good agreement with the value estimated from Eq. (\ref%
{lambda}) for the system's parameters experimentally accessible. The
relaxation and decoherence rates were assumed to be independent of the
system's parameters for simplicity. We note that the shape of the curves, in
particular the widths and the heights of the resonances, is defined by three
parameters: amplitude ($\widetilde{n}_{g}$) and the relaxation and
decoherence rates. These values can be determined from the analysis of the
widths and the heights of the resonances as e.g. in Ref. \cite{yaponci}. But
we rather fit the whole curves, which allows us to determine the system's
parameters.

\begin{figure}[h]
\includegraphics[width=4.5cm]{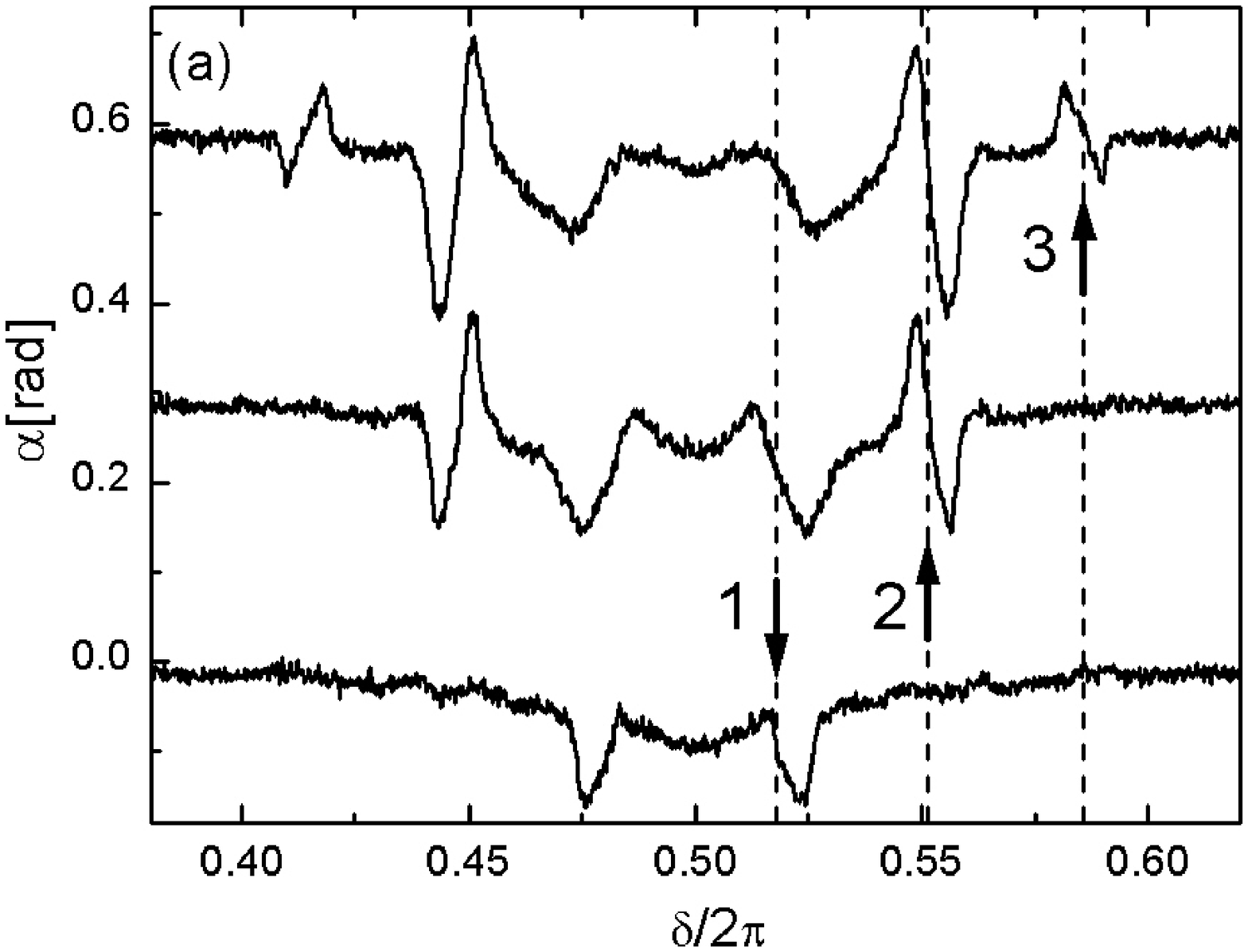}%
\includegraphics[width=4.5cm]{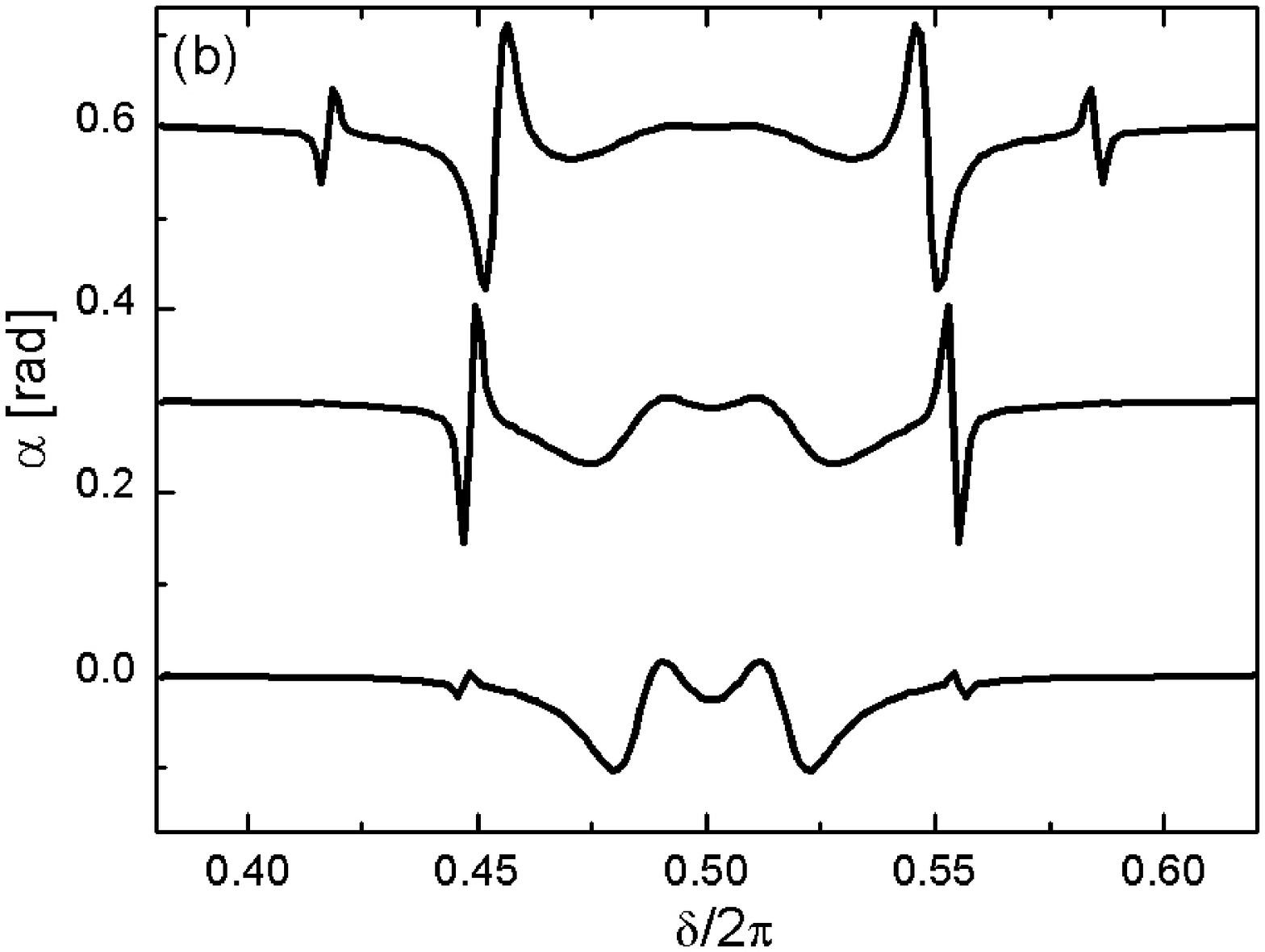}
\caption{Dependence of the tank voltage phase shift $\protect\alpha $ on the
phase difference $\protect\delta $. The curves correspond to the fixed
frequency $\protect\omega /2\protect\pi =7.05$ GHz with the varied
parameter: in experiment (a) being power of excitation (from bottom to top:
-80, -60, -57 (dB)) and in theory (b) being amplitude $\widetilde{n}_{g}$
(from bottom to top: 0.1, 0.2, 0.4). Upper curves are shifted. The arrows
show the appearance of 1-, 2-, and 3- photon excitations at $\Delta E(%
\protect\delta )=K\hbar \protect\omega $, $K=1,2,3$.}
\label{Fig5}
\end{figure}

Now we consider the dependence of the phase shift $\alpha $ on $\delta $ by
making use of Eqs. (\ref{alpha}) and (\ref{eq1})-(\ref{eq3}): see Fig. \ref%
{Fig5}. From the above considerations we have the following parameters: $%
\lambda $, $E_{C}$ and $E_{J1}-E_{J2}$. But at $\delta \neq \pi $ we also
need $E_{J1,2}$ (see Eq.(\ref{alpha})). At $\delta \neq \pi $ in the $\alpha
-\delta $-curve, due to the domination of the second term in Eq. (\ref{alpha}%
), the multiphoton resonances result in the hyperbolic-like behaviour with $%
\alpha \simeq 0$ at $\Delta E=K\hbar \omega $. We note that in the vicinity
of $\delta =\pi $ the first term in Eq. (\ref{alpha}) decreases the value of 
$\alpha $, which explains why the one-photon hyperbolic-like excitation is
not symmetric about $\alpha =0$ axis. From the position of these points,
marked with the arrows in Fig. \ref{Fig5}a (namely from the relation $\Delta
E(n_{g},\delta )=K\hbar \omega $, see Eq. (\ref{DeltaE})), we found: $%
E_{J1}/h\simeq 8E_{C}/h=40$ GHz, $E_{J2}/h\simeq 6.9E_{C}/h=34.5$ GHz, and
also $n_{g}\simeq 0.85$. With these values we have calculated the dependence
of $\alpha $ on $\delta $, shown in Fig. \ref{Fig5}b. To fit the
experimental curves we have taken: $\Gamma _{\phi }/(E_{C}/h)=0.05$ and $%
\Gamma _{relax}/(E_{C}/h)=0.03$, which correspond to the following
decoherence and relaxation times: $T_{\phi }=\Gamma _{\phi }^{-1}\simeq 4ns$
and $T_{relax}=\Gamma _{relax}^{-1}\simeq 7ns$.

\section{Conclusions}

Multiphoton (namely, one-, two-, and three-photon) excitations of the PBCQ
were observed experimentally and described theoretically. The multiphoton
transitions manifest themselves in the dependence of the tank voltage phase
shift $\alpha $ on the qubit's parameters as follows: there are resonances
in the dependence of $\alpha $ on $n_{g}$ at $\delta =\pi $ and there are
hyperbolic-like dips and peaks in the dependence of $\alpha $ on $\delta $.
Theoretical fitting has allowed us to find out the qubit's parameters,
particularly the relaxation and decoherence rates, which characterize the
decoherence processes in the system.

\begin{acknowledgments}
The authors would like to thank D. Averin and Ya. S. Greenberg for fruitful
discussions. V.I.Sh. acknowledges the financial support of the Deutsche
Forschungsgemeinschaft under grant No. KR1172/9-2. E.I. thanks the EU for
support through the RSFQubit project. A.N.O. and S.N.Sh. acknowledge the
grant \textquotedblleft Nanosystems, nanomaterials, and
nanotechnology\textquotedblright\ of the National Academy of Sciences of
Ukraine.
\end{acknowledgments}

\end{document}